\documentclass[rmp,superscriptaddress,preprint]{revtex4-1}

\usepackage{amsmath}
\usepackage{amssymb}
\usepackage{setspace}
\usepackage{graphicx}
\usepackage{natbib}
\usepackage{float}

\begin{document}
 
 %

\begin{center}
 { \large {\bf Wave Function Collapse, Non-locality,  and Space-time Structure}}


\vskip 0.3 in

{\large{\bf Tejinder P.  Singh}}

{\it Tata Institute of Fundamental Research,}
{\it Homi Bhabha Road, Mumbai 400005, India}\\
\bigskip
{\tt tpsingh@tifr.res.in}\\

\end{center}

\bigskip
\bigskip

\centerline{\bf ABSTRACT}
\bigskip
\noindent Collapse models possibly suggest the need for a better understanding of the structure of space-time.  We argue that physical space, and space-time, are emergent features of the Universe, which arise as a result of dynamical collapse of the wave-function. The starting point for this argument is the observation that classical time is external to quantum theory, and there ought to exist an equivalent reformulation which does not refer to classical time. We propose such a reformulation, based on a non-commutative special relativity. In the spirit of Trace Dynamics, the reformulation is arrived at, as a statistical thermodynamics of an underlying classical dynamics in which matter and non-commuting space-time degrees of freedom are matrices obeying arbitrary commutation relations. Inevitable statistical fluctuations around equilibrium can explain the emergence of classical matter fields and classical space-time, in the limit in which the universe is dominated by macroscopic objects. The underlying non-commutative structure of space-time also helps understand better the peculiar nature of quantum non-locality, where the effect of wave-function collapse in entangled systems is felt across space-like separations.

\noindent 

\vskip 1 in

\setstretch{1.4}

\section{Introduction}

\noindent Dynamical collapse models explain the collapse of the wave-function and provide a solution of the quantum measurement problem, by proposing that the Schr\"{o}dinger equation is approximate. It is an approximation to a stochastic nonlinear dynamics, with the stochastic nonlinear aspect becoming more and more important as one progresses from microscopic systems to macroscopic ones. They are the only concrete class of models whose predictions differ from those of quantum theory in the mesoscopic and macroscopic domain. These predictions are currently being tested in a variety of experiments which are putting ever tighter constraints on departure from the standard linear theory \cite{RMP:2012}. Yet, a large part of the parameter space remains to be ruled out. In Section II we give a very brief description of these models and then provide an overview of the status of their experimental tests.

The Schr\"odinger equation is of course non-relativistic, and so are the collapse models described in Section II.  According to theory, collapse is an instantaneous process, and experiments also confirm that collapse of the wave-function at one location in space  produces an influence at another space-like separated location. This suggests that making a relativistic theory of collapse will be challenging, and indeed this turns out to be so. In Section II we briefly describe the current status of various attempts to make a relativistic theory of collapse. 

It is possible that sometime in the future we will have a convincing relativistic theory of collapse of the wave function. In the present work though, we attempt to make a case that there is a fundamental reason why collapse and relativity are incompatible. The reason is what we call the `problem of time' in quantum theory. As we argue in Section III, time being a classical concept, it is external to quantum theory. There should exist an equivalent reformulation of quantum theory which does not refer to an external classical time.    Such a reformulation suggests spacetime is fundamentally not classical, but perhaps has a noncommutative structure. In the absence of collapse, as we shall argue, there is a map from non-commutative space-time to ordinary space-time, so that linear quantum theory is compatible with relativity.

However when collapse is included, this map from non-commutative to ordinary space-time is not possible, and relativistic collapse can be described meaningfully only in a non-commutative space-time. This explains why collapse appears instantaneous and non-local, when seen from the [inadmissible] vantage point of ordinary space-time. We wish to suggest that the apparently acausal nature of collapse is incompatible with ordinary relativity, but compatible with the non-commutative space-time implied by a resolution of the problem of time in quantum theory. This aspect is discussed in Section IV.

Our work described in Sections III and IV represents a partial mathematical development of the proposed physical ideas. Considerable more work remains to be done in order to arrive at a complete mathematical model.
  
 \section{Collapse Models and their Experimental Tests}   
The most advanced dynamical collapse model is known as Continuous Spontaneous 
Localization (CSL)  \cite{Bassi:03,Pearle:89,Ghirardi:86,Ghirardi2:90} . Its `mass proportional' version is described by the following stochastic nonlinear modification of the Schr\"odinger equation, in Fock space:
\begin{eqnarray} \label{eq:csl-massa}
d\psi_t =   \left[-\frac{i}{\hbar}Hdt \right. 
+ \frac{\sqrt{\gamma}}{m_{0}}\int d\mathbf{x} (M(\mathbf{x}) - \langle M(\mathbf{x}) \rangle_t)
dW_{t}(\mathbf{x}) \nonumber 
 -  \left. \frac{\gamma}{2m_{0}^{2}} \int d\mathbf{x}\,
(M(\mathbf{x}) - \langle M(\mathbf{x}) \rangle_t)^2 dt\right] \psi_t
\end{eqnarray}
The first term describes standard Schr\"{o}dinger evolution, with $H$ being the quantum Hamiltonian. The second and third terms are the new terms which induce dynamical collapse. We note that the new terms are non-unitary, yet they maintain the norm-preserving nature of the evolution.  $\gamma$ is a (positive)  new  constant  of nature which determines the strength of collapse - it must be detected in experiments, if CSL is to be confirmed. $m_0$ is a reference mass, conventionally chosen to be the mass of the nucleon. $M({\bf x})$ is the mass density operator, defined as
\begin{eqnarray}
M(\mathbf{x})
& = & \underset{j}{\sum}m_{j}N_{j}(\mathbf{x}), \label{eq:dsfjdhz}\\
N_{j}(\mathbf{x})
& = & \int d\mathbf{y}g(\mathbf{y-x})
\psi_{j}^{\dagger}(\mathbf{y})\psi_{j}(\mathbf{y}), \qquad
\end{eqnarray}
Here $\psi_{j}^{\dagger}(\mathbf{y})$ and
$\psi_{j}(\mathbf{y})$ are the  creation and
annihilation operators, respectively,  for a particle $j$ at the location
$\mathbf{y}$. The smearing function $g({\bf x})$ is given by
\begin{equation} \label{eq:nnbnm}
g(\mathbf{x}) \; = \; \frac{1}{\left(\sqrt{2\pi}r_{\text{\tiny C}}\right)^{3}}\;
e^{-\mathbf{x}^{2}/2r_{\text{\tiny C}}^{2}},
\end{equation}
where $r_{\text{\tiny C}}$ is the second new phenomenological constant of the model which also should be detected experimentally. This is the length scale to which the collapsed wave function is localised. Stochasticity in the CSL model is described by $W_{t}\left(\mathbf{x}\right)$ which is an
ensemble of independent Wiener processes, one for each point in space. 

The dynamical collapse of the wave function is understood in a straightforward manner by constructing the master equation for the density matrix, where the CSL terms cause decay of the off-diagonal elements. The decay constant is given by
\begin{equation}
\lambda_{\text{\tiny CSL}} \; = \; \frac{\gamma}{(4\pi r_{\text{\tiny C}}^2)^{3/2}}.
\end{equation}
The CSL model assumes the value
\begin{equation}
\lambda_{\text{\tiny CSL}} \approx  10^{-17} \text{s}^{-1}.
\label{lambda}
\end{equation} 
This is the most conservative [i.e. smallest] value consistent with phenomenology - it gives a sufficiently high value for the superposition lifetime of a nucleon in two different position states, and causes macroscopic superpositions to collapse sufficiently fast. The model proposes a value of about $10^{-5}$ cm for $r_{\text{\tiny C}}$ - values much smaller or much larger than this would contradict observations.

Collapse models are phenomenological, designed for the express purpose of solving the measurement problem and explaining the Born probability rule, and explaining why macroscopic objects behave classically. Essentially, because of the nonlinearity, the superposition principle becomes approximate: the lifetime of a quantum superposition [infinite according to the linear theory] is now finite, becoming increasingly smaller with increasing mass. Experiments test CSL by looking for this spontaneous breakdown of superposition, and since violation of superposition has not been observed, this puts bounds on $\lambda_{\text{\tiny CSL}}$ and on $r_{\text{\tiny C}}$ - essentially the allowed region in the $\lambda_{\text{\tiny CSL}} - r_{\text{\tiny C}}$ plane gets restricted.

The presence of the stochastic noise field in the theory results in a tiny violation of energy-momentum conservation.  This leads to an energy gain and heating, which for a particle of mass $M$, is given by the rate \cite{Pearle:94,Adler3:07}
\begin{equation}
\frac{dE}{dt} = \frac{3\lambda_{\text{\tiny CSL}}}{4}\frac{\hbar^2}{r_{\text{\tiny C}}^2} \frac{M}{m_N^2}.
\label{heating}
\end{equation}
A class of experiments put important bounds on CSL parameters by looking for this anomalous heating in controlled systems.

\subsection{Experimental Tests} 
Collapse models such as CSL make predictions for experiments; these are different from the predictions made by quantum theory, because of the introduction of the new non-linear stochastic terms, and the two new constants of nature, $\lambda_{CSL}$ and $r_C$. The most direct consequence is that the principle of quantum linear superposition is not an exact principle of nature, but an approximate one. What this means is that a quantum superposition of states of a microscopic system such as an electron lasts for an astronomically long time, comparable to the age of the Universe. However, the quantum superposition of states of a macroscopic system lasts for an extremely short time interval, and this is what provides a resolution of the quantum measurement problem, and explains the quantum-classical transition. Thus if one does a matter-wave interferometry experiment with progressively larger objects, going from electrons to neutrons, atoms, molecules and ever larger objects, then the fact that one continues to observe the interference pattern (which of course is a consequence of superposition being valid) puts an upper bound on $\lambda_{CSL}$. Absence of the interference pattern would confirm CSL and the value of $\lambda_{CSL}$, provided all other known sources of decoherence have been ruled out [such as collisional and thermal decoherence].  The largest molecule for which an interference experiment has been done to date has a mass of about $10^{4}$ amu, which implies an upper bound $\lambda_{CSL}\sim 10^{-5}$, assuming the plausible value of $r_C\sim 10^{-5}$ cm. It has been suggested that if interferometry experiments can be carried out for a mass all the way upto $10^{9}$ amu, and interference continues to be observed, that will push the  upper bound on $\lambda_{CSL}$ all the way down to (\ref{lambda}) and essentially rule out the spontaneous collapse model. The state of the art in these experiments, and their technological challenges and future prospects are discussed for instance in \cite{RMP:2012,Arndt:2014, Ulbricht:2016}.

Optomechanical experiments of massive objects, typically in the nano-range and with a mass of about $10^{15}$ amu, aim to create position superpositions of mesoscopic objects by entangling them with light. A challenge for these experiments is that the position separation they achieve in the superposition is much smaller than the favoured value of $10^{-5}$ cm for $r_C$ \cite{Aspelmeyer:2014}.

Very promising progress is currently being made on testing CSL through non-interferometric tests, using the heating effect described by Eqn. (\ref{heating}). The idea being that if one observes an isolated system for a certain length of time, then the CSL heating effect will leave an imprint, such as an anomalous rise in temperature, or an anomalous random walk of an isolated nanosphere, or an anomalous noise for which there is no other plausible explanation. There are various ways in this effect can be used to put bounds on CSL. Firstly, the heating would contribute to results of laboratory experiments and astronomical observations that have already been carried out and which are in agreement with the predictions of quantum theory. In order that CSL should not disagree with the already observed result, one puts an upper bound on $\lambda_{CSL}$. These bounds come from ionization of the intergalactic medium, decay of supercurrents, excitation of bound atomic and nuclear systems, absence of proton decay, heating of ultra-cold atoms in Bose-Einstein condensates, and rate of spontaneous 11 keV emission from Germanium \cite{Adler3:07,Laloe:2014}. This last effect provides the strongest bound to date on the CSL parameter: $\lambda_{CSL}<10^{-11}$. Very impressive bounds have recently been put from the knowledge of the thermal noise in LIGO and the highly sensitive LISA pathfinder.  Proposed tests of anomalous heating include observing the random walk of a micro-object at low temperatures, the anomalous motion of a levitated nanosphere, and spectral broadening. \cite{PEARLE5, Bera:2015,Rashid:2016,Bahrami:2014,Barker:2015,Bahrami:2014a,Goldwater:2015,Bilardello:2016,Bedingham:2016}. The current bounds on $\lambda_{CSL}$ and $r_C$ can be found in Fig. 2 of \cite{Carlesso:2016}.

A landmark recent experiment reports the strongest direct upper bound obtained by measuring the thermal noise in a ultra-cold nano-cantilever cooled to milli-kelvin temperatures \cite{Vinante:2016a}. [For the exclusion plot in the 
$\lambda-r_C$ plane see Fig. 3 of this paper]. A more refined version of the experiment reports the first  possible observation and detection of a `non-thermal force noise of unknown origin' which is compatible with previously known bounds on CSL \cite{Vinante:2016b}.. This is perhaps the first time ever that a possible departure from quantum theory has been reported. It remains to be seen if the result will survive further similar tests of CSL.

\subsection{Relativistic Collapse Models} 
A major remaining challenge is to construct a satisfactory relativistic quantum field theory of spontaneous collapse. Many serious attempts have been made in this direction \cite{Ghirardi:90,Pearle:90,Nicrosini:03,Myrvold:09,Pearle2:99,Bedingham:10,Bedingham:11,Tumulka2:06,Dowker:04,Dowker2:04,Bedingham:2016a}. The first such attempt was to convert the CSL equation into a Tomanaga-Schwinger type generalisation, with evolution being described with respect to an arbitrary space-like hypersurface. The problem arises in choosing a Lorentz invariant stochastic 
noise which couples locally to the quantum fields - typically such a choice leads to divergences, in the form 
of an infinite rate of energy production. Changing the coupling to a non-local coupling creates other difficulties. Various attempts have been made, and work is still in progress, to resolve the problem posed by divergences. Some of the other articles in this volume discuss some of these developments in greater detail. In our present article, we take the stance that there are fundamental reasons [problem of time, quantum non-locality] because of which we possibly cannot have a relativistic generalisaton of CSL and a revised picture of spacetime structure is essential, to make collapse consistent with relativity. We attempt to make the case that collapse is consistent with a non-commutative generalisation of special relativity, but not with ordinary special relativity.

\section{The problem of time in quantum theory} 
As in most other physical systems, evolution in time is central to the understanding of quantum systems. The time that is used to define evolution in quantum theory is obviously part of the classical space-time manifold, and the space-time geometry could be either Minkowski,  or curved - in the latter case there is a non-trivial metric overlying the manifold. One could take these classical aspects [manifold and geometry] as external givens in a quantum theory. However, doing so leads to a problem of principle, which suggests that the present formulation of quantum theory is incomplete.

According to the Einstein hole argument, the physical metric is required, in order to give an operational meaning to the space-time manifold. The metric is determined, according to the laws of general relativity, by the classical distribution of macroscopic material bodies and matter fields. But these material bodies are a limiting case of quantum theory. Thus through its need for time, whose operational definition requires a space-time metric and hence the presence of classical objects, the formulation of quantum theory depends on its own classical limit [self-reference]. From a fundamental viewpoint, this cannot be considered satisfactory. The classical-quantum divide reminds of the Copenhagen interpretation, which we know has its shortcomings.

 In particular, we could envisage a universe, or an epoch in the history of the universe [such as soon after the big bang] where classical material bodies are absent and matter is entirely microscopic in nature. Such matter fields would not give rise to a classical metric [unless one asserts as a matter of principle that gravity {\it is} classical no matter what, and/or one invokes a semiclassical theory of gravity, which by itself is problematic). Gravity produced by microscopic matter fields will possess fluctuations, as a result of which one can no longer give physical meaning to the underlying space-time manifold.

We may conclude from the reasoning in the previous two paragraphs that there ought to exist an equivalent reformulation of quantum theory which does not refer to classical time. The reformulation would reduce to ordinary quantum theory as and when the universe is dominated by classical macroscopic objects, so that a classical space-time manifold maybe meaningfully defined \cite{Singh:2006}.

In order to arrive at such a reformulation, we ask as to what kind of space-time geometry might be compatible with quantum theory. The choices are undoubtedly vast, but as a starting point we propose that in analogy with the generic non-commutative nature of  observables in quantum theory, space-time coordinates do not commute with each other in the sought for reformulation. This seems a reasonable demand, if quantum space-time is to be produced by quantum fields, and if the canonical degrees of freedom in the latter have a non-commutative structure. We are thus led to a non-commutative geometry, with its own metric. We try to arrive at a model by suggesting that the reformulation should be invariant under general coordinate transformations of non-commuting coordinates, thus taking a clue from the principle of general covariance.

Ideally, a successful reformulation should also address those other aspects of quantum theory which maybe considered its shortcomings. We have in mind two aspects in particular. The first is that the theory is constructed, not from first principles, but by `quantizing' its own classical limit. Thus for instance Poisson brackets from the classical theory are replaced by quantum canonical commutation relations, essentially as an ad hoc recipe. A more fundamental description of quantum theory would motivate the commutation relations in a manner independent of its classical limit. The second has to do with the phenomenological nature of collapse models: the origin of the noise field must ultimately be understood at a fundamental level. Here it is noteworthy that the standard quantum theory part of the collapse model is noise-free, and the fluctuations serve to modify the quantum evolution (negligibly in the micro-limit and significantly in the macro-world). This division of the collapse dynamics between a noise-free part and a noisy part is suggestive of statistical thermodynamics; as if the phenomenological collapse model is the thermodynamic limit of an underlying microscopic theory.  From the underlying theory one may derive, after statistical averaging, quantum theory as the state of thermodynamic equilibrium; and the noise part is Brownian motion fluctuations around equilibrium. The theory of Trace Dynamics, developed by Adler and collaborators, goes a long way in providing such an underlying theory. We have made some preliminary attempts in generalizing the ideas of Trace Dynamics to achieve  a reformulation of quantum theory without classical time. Much remains to be done, and at this stage we cannot claim that we have necessarily found the right direction for our purpose.

This is not the appropriate place for describing Trace Dynamics [TD] in detail - adequate details are available elsewhere \cite{Adler:04,RMP:2012,Adler:94,Adler-Millard:1996}. In brief, TD is the classical dynamics of matrices that exist on a background space-time, and whose elements are complex Grassmann numbers. The matrices could be `bosonic (B) / fermionic (F)', i.e. their elements are even-grade / odd-grade elements of the Grassmann algebra, respectively. One can construct a polynomial $P$ from these matrices, obtain its trace ${\bf P} = Tr P$, and define a so-called Trace derivative of ${\bf P}$ with respect to a matrix. Given this, a trace Lagrangian can be constructed  from the matrices (i.e. operators) $\{q_r\}$ and their time derivatives $\{\dot{q}_r\}$. Then one can develop a classical Lagrangian and Hamiltonian dynamics for the system of matrices, in the conventional way. The configuration variables and their canonical momenta all possess arbitrary commutation relations with each other. However, as a result of the global unitary invariance of the  trace Hamiltonian, the theory possesses a remarkable conserved charge made up from the commutators and anti-commutators.
\begin{equation}
\tilde{C} = \sum_B [q_r,p_r] -\sum_F \{q_r,p_r\} 
\end{equation}
This charge, which has the dimensions of action, and which we call the Adler-Millard charge, is what makes TD uniquely different from the classical mechanics of point particles. It plays a key role in the emergence of quantum theory from TD. Assuming that one wishes to observe the dynamics at a coarse-grained level, one constructs the statistical mechanics of TD in the usual way, by defining a canonical ensemble, and then determining the equilibrium state by maximising the entropy, subject to conservation laws. The canonical average of the Adler-Millard charge takes the form
\begin{equation}
\langle \tilde{C} \rangle _{AV}  = i_{eff}\hbar; \qquad i_{eff} = diag(i,-i,i,-i,...)
\label{canam}
\end{equation}
where $\hbar$ is a real positive constant with dimensions of action, which is eventually identified with Planck's constant. 

Having found the equilibrium state, one can draw important conclusions from it. A general Ward identity [analog of the equipartition theorem in statistical mechanics] is derived as a consequence of invariance of canonical averages under constant shifts in phase space. Its implications are closely connected with the existence of the Adler-Millard charge and its canonical average   (\ref{canam}). After applying certain realistic assumptions, the canonical averages of the TD operators are shown to obey Heisenberg equations of motion, and the canonical commutation relations of quantum theory. In this sense, upon the identification of  canonical averages of TD with Wightman functions in quantum field theory, one finds quantum theory emergent as a thermodynamic approximation to TD. The passage from the Heisenberg picture to the Schr\"{o}dinger picture is made in the standard manner, to arrive at the non-relativistic Schr\"{o}dinger equation.

Next, one takes account of the ever-present thermal fluctuations around equilibrium; in particular there are fluctuations in the Adler-Millard charge about its canonical value (\ref{canam}). As a result, the Schr\"{o}dinger equation picks up [linear] stochastic correction terms. If one invokes the assumptions that evolution should be norm-preserving in spite of the stochastic corrections, and that there should be no superluminal signalling, the modified Schr\"{o}dinger equation becomes non-linear and has the generic structure of a spontaneous collapse model. One can demonstrate dynamic collapse which obeys the Born probability rule. Of course the theory is not well-developed enough at this stage to uniquely pick out the CSL model, or to  predict the numerical values of the constants $\lambda_{CSL}$ and $r_{\text{\tiny C}}$. Moreover, the assumption of norm-preservation in the presence of stochastic corrections is ad hoc. Norm-preservation should follow from a deeper principle, yet to be discovered. Another noteworthy feature is that in order to define a thermal equilibrium state, one is compelled to pick out a special frame of reference, possibly the cosmological rest frame of the cosmic microwave background. 

In spite of important issues which remain to be resolved, Trace Dynamics is a significant example of an underlying theory from which quantum theory and collapse models are emergent. We have attempted to extend the ideas of TD to arrive at a reformulation of quantum theory without classical time, by raising time and space coordinates to the level of non-commuting matrices, as is done in TD for the canonical degrees of freedom. This program has met with partial success, in that it could be implemented for Minkowski 
space-time. The case for a curved space-time remains to be developed, although a heuristic outline is available \cite{Singh:2012}.

On a non-commutative Minkowski space-time, with $(\hat{t},\hat{x},\hat{y},\hat{z})$ as non-commuting operators having arbitrary commutation relations, we define a trace proper time as follows \cite{Lochan-Singh:2011}
\begin{equation}
ds^2 = Tr d\hat{s}^2 \equiv Tr [ d\hat{t}^2 - d\hat{x}^2 - d\hat{y}^2 - d \hat{z}^2 ] 
\label{nsr}
\end{equation}
This line-element can be shown to be invariant under Lorentz transformations. Matter degrees of freedom `live' on this non-commutative space-time, and one can define a Poincar\`e invariant dynamics, by first introducing a four-vector $\hat{x}^{\mu} = (\hat{t},{\bf x})$ and defining four velocity as $\hat{u}^{\mu}=d\hat{x}^{\mu}/ds$. Lagrangian and Hamiltonian dynamics can then be constructed in the spirit of Trace Dynamics, using the trace proper time to define evolution. As before, there exists in the theory, as a consequence of global unitary invariance, a conserved Adler-Millard charge for the matter degrees of freedom $\hat{y}^\mu$
\begin{equation}
\hat{Q} = \sum_{r\in B} [\hat{y}_r,\hat{p}_r] - \sum_{r\in F} \{\hat{y}_r,\hat{p}_r\}
\end{equation}
The important generalization is that, there is associated, with every degree of freedom, apart from the canonical pair $(\hat{q}, \hat{p})$, a conjugate pair $(\hat{E},\hat{t})$,  where $\hat{E}$ is the energy operator conjugate to $\hat{t}$. The metric above, although it is Lorentz invariant, does not admit a light-cone structure, or point structure of ordinary space-time. This is what allows the recovery of a quantum theory without classical time. 

From this point on, the construction parallels that in TD, namely an equilibrium statistical thermodynamics of the underlying classical theory is constructed, with coordinate time $\hat{t}$ now an operator, and evolution being described with  respect to trace proper time ${s}$. The quantum commutators emerge at the thermodynamic level, with the added feature that there is now an energy-time commutator as well. In the non-relativistic limit one obtains the generalised Schr\"{o}dinger equation
\begin{equation}
i\hbar \frac{d\Psi}{ds} = H\Psi(s)
\label{gqd}
\end{equation}
where the configuration variables now include also the time operator $\hat{t}$. It is important to emphasize that the configuration variables commute with each other. We call this a Generalised Quantum Dynamics (GQD), in which there is no classical space-time background, and all matter and space-time degrees of freedom have operator status. This is the sought for reformulation of quantum theory which does not refer to classical time \cite{Lochan:2012}.

We now provide a heuristic picture as to how standard quantum theory on a classical space-time background is recovered from here \cite{Singh:2012}. For that, we need to first understand how GQD can give rise to a classical space-time. As we argued above, we expect a classical space-time to exist only when the universe is dominated by classical macroscopic objects.  Consider now the role of statistical fluctuations of the Adler-Millard charge about equilibrium, which modify the Schr\"{o}dinger equation into a non-linear equation, as discussed above. And we consider the epoch of the very early universe, when tiny primordial matter density perturbations are present, which subsequently grow into large scale structure. As these perturbations grow in magnitude, the associated statistical fluctuations of the non-linear stochastic Schr\"{o}dinger equation become more and more significant, thereby eventually bringing about the classicalization of these density perturbations, in the spirit of collapse models. The localization takes place not only in space, but also in time! The time operator associated with every object becomes classical, by which we mean that it takes the form of a $c$-number times a unit matrix.
Moreover, the statistical fluctuations associated with the operator space-time degrees of freedom also become more and more significant, leading to the emergence of a classical space-time. 

We conclude that the localization of macroscopic objects occurs in conjunction with the emergence of a classical space-time. This is consistent with the Einstein hole argument, namely that classical matter fields and the gravitational metric which they  produce are both required in order to give physical meaning to the point structure of space-time.  Only when the Universe is dominated by macroscopic objects, as is true for today's Universe, can we assume the existence of a classical space-time. In this approximation, the trace proper time $s$ of the generalised TD can be identified with ordinary classical proper time. Once the Universe achieves such a  classical state, it sustains itself therein, because of continual action of stochastic fluctuations on macroscopic objects, thus concurrently achieving  the existence of a classical space-time geometry. Because the underlying generalized trace dynamics is Lorentz invariant, the emergent classical  space-time is also locally Lorentz invariant. There is however a fundamental  difference: in contrast with the underlying theory, in the classical approximation,  light-cone structure, and causality, are approximate emergent features, since the space-time coordinates are now $c$-numbers. 

We may now argue how standard quantum theory for a microscopic system emerges from GQD, once a classical background universe with its classical space-time, is given. Regardless of whether or not there is a classical space-time background, a microscopic system  is described at a fundamental level via its non-commutative space-time (\ref{nsr}), through the associated generalized TD. Subsequent to coarse-graining, and construction of the equilibrium statistical thermodynamics, this leads to the system's GQD (\ref{gqd}) with its own trace time. In the approximation   that the stochastic fluctuations can be ignored, this GQD possesses commuting $\hat t$ and $\hat {\bf x}$ operators. As a consequence of their commutativity, these can be mapped to the $c$-number $t$ and ${\bf x}$ coordinates of the pre-existing classical universe, and trace time can then be mapped to ordinary proper time. This is hence a mapping to ordinary space-time, and one recovers standard  quantum mechanics in this manner. If this program can be constructed rigorously, it will  explain as to how ordinary quantum theory is recovered from a reformulation which does not depend on classical time. Statistical fluctuations, and collapse models, then explain the classical limit of quantum mechanics, as before, through the stochastic non-linear Schr\"{o}dinger equation. 

We regard it as a profound inference that the dynamical collapse of the wave-function is intimately and directly responsible for the emergence of a classical space-time. Collapse of the wave-function is not just an aspect concerning the interaction of a quantum system with a measuring apparatus in the laboratory. Rather, wave-function collapse is responsible for the emergence of classical objects in the universe, and is hence also responsible for the emergence of classical space-time. This appears to be a robust and intuitive inference, independent of the mathematical details of trace dynamics. It also appears plausible because it is collapse which localizes macroscopic objects, thereby creating space as that is `between objects'. It does not appear meaningful to talk of space if there were no collapse and macroscopic objects were delocalised and `everything was everywhere' so to say: what use would the concept of space be of, in that case? This possibly helps us understand another strange feature of collapse: the wave-function, satisfying the Schr\"{o}dinger equation, lives in a Hilbert space, not in physical space. On the other hand, the so-called jump operator of collapse models  maps the wave-function to a localized position in physical space: there is somehow a mysterious connection between the complex wave-function living in Hilbert space, and the real point in physical space to which the particle being described by the wave-function gets localised. This apparent mystery can be removed if we think of the collapse process as taking place in the non-commutative space of the underlying trace dynamics. The process is then described in the Hilbert space of the states in the underlying generalised trace dynamics, which includes the non-commutative space, to which the physical space is an emergent approximation. The state of the system, fundamentally speaking, does not jump from the Hilbert space to physical space; rather it always lives in the non-commutative space, before the jump, as well as after the jump. 

As an illustration, let us consider the famous double slit experiment with electrons. Electrons are released one by one from the electron gun; each electron passes through the slits, and upon reaching the screen collapses to one or the other position, in accordance with the Born probability rule. Once a large number of electrons have collected on the screen, they are distributed as per the Born rule, forming the interference pattern. What happens to an electron after it leaves the gun, and before it reaches the screen? There is really no satisfactory description of `what happens',  in standard quantum theory. The electron is supposed to behave like a wave and go through both slits, and the two wavelets (one from each screen) interfere with each other. But what is this wave a wave of? It cannot possibly be the wave-function - which is a complex valued quantity, which lives in Hilbert space, not in physical space. Often it is suggested that the wave is a probability wave - that also seems strange; because then we do not seem to attach any physical reality to the wave-function. Comprehending the situation becomes easier if we think of this entire process as taking place in the underlying non-commutative space, to which ordinary space is an approximation. The complex-valued description of the state as a `wave' appears natural if we think of it as a wave in `non-commutative' space. Problem arises only when we try to describe quantum phenomena on a classical space-time background - the two are in conflict with each other; classical space-time is external to quantum theory, and not really a part of a fundamental description of quantum theory. 

We conclude that in our study, the problem of time and the problem of measurement are closely related. Suppose we start from a formulation of quantum theory which does not refer to classical time. Then, so as to recover classical time and space-time geometry from this reformulation, we also need to recover the macroscopic limit of matter fields. This is because classical geometry and classical matter fields co-exist. Furthermore,  to explain the classical behaviour of macroscopic objects is the same as solving the quantum measurement problem. This is because the latter problem can be stated differently as follows: why are macroscopic objects never observed in superposition of position states?  The measurement problem is a small part of a larger problem: how does the classical structure of space-time and matter degrees of freedom arise from an underlying quantum theory of matter and space-time? We have not yet addressed the issue of the gravitational field on non-commutative space-time, nor whether one is constrained to select a specific frame of reference when studying statistical fluctuations and collapse, in GQD. 

\section{Quantum non-locality and space-time structure} 
Non-local quantum correlations cannot be used for superluminal signalling; nonetheless they suggest an acausal influence amongst the entangled particle pairs, when a measurement is made on one of them. It is as if the pair behaves like a rigid body, even if the two particles lie outside each other's light cone. To many researchers, this calls for the need for an explanation, even though there is no known experimental conflict between quantum theory and special relativity. Since the influence arises only when a measurement is made and hence wave-function collapse takes place, it is natural to expect that an explanation for the non-local influence must be tied up with the explanation for wave-function collapse. 

A possible explanation has been suggested by Adler, in the context of trace dynamics and the related explanation for wave-function collapse in terms of the statistical fluctuations of the Adler-Millard charge: these modify the Schr\"{o}dinger equation to a stochastic non-linear equation. Since the stochastic terms are non-linear, they do result in violation of local causality, as changes in the wave-function at one point in space are instantaneously communicated to every other point. However, the average of the stochastic terms over the density matrix obeys a linear equation, thus rendering superluminal signalling impossible.  Adler also notes that when the statistical fluctuations are taken into account (i.e. when one goes beyond the linear quantum theory) it becomes necessary to choose a specific inertial frame of reference (so as to meaningfully define the canonical ensemble). This choice of frame, which is possibly the cosmological rest frame of the CMB, breaks Lorentz invariance and collapse is hence not a relativistically invariant process. This could be one possible reason why it is difficult to make a relativistic theory of collapse.

While at some level this maybe an adequate explanation, one might still be left wondering if `instantaneous communication' across vast expanses of space can be regarded as physically reasonable. There is also the earlier question of the complex wave-function living in Hilbert space, and how  it could be responsible for such communication in physical space. Perhaps the existence of such communication might appear more reasonable if there was a quantum  notion of `zero physical distance' as opposed to the enormous spatial distance that we perceive in ordinary space, over which the entangled pair `communicates'. In the context of the scheme developed in the present work, we offer such an explanation as follows \cite{Banerjee:2016}.

We explained above how a classical space-time emerges from the GQD, in conjunction with the domination of the universe by macroscopic objects. Given this background universe, let us consider the entangled EPR pair of particles. So long as the pair is in  flight, its evolution is described by the Schr\"{o}dinger equation, which is equivalent to the GQD given by Eqn. (\ref{gqd}). This equivalence is possible because, in the absence of statistical fluctuations (which become significant only at the onset of collapse), the commuting operator coordinates ($\hat{t}, \hat{\bf x})$ can be mapped to ordinary space-time cootdinates
(${t}, {\bf x})$.
However, when one of the entangled particles in the pair interacts with the classical apparatus, and the measurement is done, the collapse inducing stochastic fluctuations  of the Adler-Millard charge become significant. By implication, fluctuations in the space-time operators $\hat{t}, \hat{\bf x}$ associated with the quantum system become important. These operators hence carry information about the arbitrary commutation relations of the  generalized TD and therefore they no longer commute with each other. This implies that these operators cannot be mapped to the space-time coordinates $(t,{\bf x})$ of special relativity.   As a result, one can define simultaneity only with respect to the trace time $s$, and there is no relativistic theory of wave function collapse, because it is not possible to describe the collapse process on a classical space-time background.

 In this scenario, collapse and the non-local quantum correlation   takes place only in the non-commutative space-time (\ref{nsr}), which is devoid of point structure,  light-cone structure, and also devoid of the notion of spatial distance. Therefore we can only say that collapse and the so-called influence from one particle to the other takes place at a specific trace time, which is Lorentz invariant, and it is not physically meaningful to talk of an instantaneous influence travelling across physical space, nor should  we call the correlation non-local. In this scenario  the entangled pair does not know distance. The state can be meaningfully described only in the non-commutative space-time, as in the case of the double-slit experiment. We once again conclude that removing classical space-time from quantum theory removes another  of its peculiarities, i.e. the so-called spooky action at a distance.

If we attempt to view and describe the quantum measurement on the entangled quantum pair from the view-point of the Minkowski space-time of ordinary special relativity, then the process  appears to violate local causality.  However, such a description cannot be considered valid, because there does not exist a map from the fluctuating and non-commuting 
(${\hat t}, \hat{\bf x}$) operators to the commuting $t$ and ${\bf x}$ space-time coordinates of ordinary special relativity. No such map exists even in the non-relativistic case. However, in the non-relativistic case,  because there exists  an absolute time, it is possible to  model the statistical fluctuations as a stochastic field on a given space-time background. This  is what is done in collapse models: collapse is instantaneous in this absolute time but it does not violate local causality.

\bigskip

\noindent {\bf Conclusions}: Dynamical collapse models are phenomenological, and there ought to exist an underlying theory from which they are emergent. The spooky action at a distance in non-local quantum correlations possibly suggests a need for revising our understanding of space-time structure. We have argued that resolving the problem of time in quantum theory suggests that space and space-time are emergent concepts, resulting from dynamical collapse of the wave-function of macroscopic objects in the universe. The underlying non-commutative structure of space-time makes it much easier to understand the apparently non-local influence in EPR correlations. We are suggesting that many of the apparently puzzling features of quantum theory arise because we attempt to describe the theory on a classical space-time background. This choice of background is unsatisfactory and inadequate for describing certain quantum phenomena, because it is approximate and external to quantum theory. We hope that further work will put these heuristic ideas on a firm mathematical footing, and also suggest experiments to test these ideas.

\bigskip
\noindent {\bf Acknowledgement}: I have benefitted greatly from useful discussions with my collaborators Angelo Bassi, Hendrik Ulbricht, Saikat Ghosh, Shreya Banerjee, Srimanta Banerjee, Sayantani Bera, Suratna Das, Sandro Donadi, Suman Ghosh, Kinjalk Lochan, Seema Satin, Priyanka Giri, Navya Gupta, Bhawna Motwani, Ravi Mohan and Anushrut Sharma.

 \bigskip


\bibliography{biblioqmtstorsion}

\def\polhk#1{\setbox0=\hbox{#1}{\ooalign{\hidewidth
  \lower1.5ex\hbox{`}\hidewidth\crcr\unhbox0}}} \def\cprime{$'$}
  \def\cprime{$'$}
\begin{thebibliography}{42}%
\makeatletter
\providecommand \@ifxundefined [1]{%
 \@ifx{#1\undefined}
}%
\providecommand \@ifnum [1]{%
 \ifnum #1\expandafter \@firstoftwo
 \else \expandafter \@secondoftwo
 \fi
}%
\providecommand \@ifx [1]{%
 \ifx #1\expandafter \@firstoftwo
 \else \expandafter \@secondoftwo
 \fi
}%
\providecommand \natexlab [1]{#1}%
\providecommand \enquote  [1]{``#1''}%
\providecommand \bibnamefont  [1]{#1}%
\providecommand \bibfnamefont [1]{#1}%
\providecommand \citenamefont [1]{#1}%
\providecommand \href@noop [0]{\@secondoftwo}%
\providecommand \href [0]{\begingroup \@sanitize@url \@href}%
\providecommand \@href[1]{\@@startlink{#1}\@@href}%
\providecommand \@@href[1]{\endgroup#1\@@endlink}%
\providecommand \@sanitize@url [0]{\catcode `\\12\catcode `\$12\catcode
  `\&12\catcode `\#12\catcode `\^12\catcode `\_12\catcode `\%12\relax}%
\providecommand \@@startlink[1]{}%
\providecommand \@@endlink[0]{}%
\providecommand \url  [0]{\begingroup\@sanitize@url \@url }%
\providecommand \@url [1]{\endgroup\@href {#1}{\urlprefix }}%
\providecommand \urlprefix  [0]{URL }%
\providecommand \Eprint [0]{\href }%
\providecommand \doibase [0]{http://dx.doi.org/}%
\providecommand \selectlanguage [0]{\@gobble}%
\providecommand \bibinfo  [0]{\@secondoftwo}%
\providecommand \bibfield  [0]{\@secondoftwo}%
\providecommand \translation [1]{[#1]}%
\providecommand \BibitemOpen [0]{}%
\providecommand \bibitemStop [0]{}%
\providecommand \bibitemNoStop [0]{.\EOS\space}%
\providecommand \EOS [0]{\spacefactor3000\relax}%
\providecommand \BibitemShut  [1]{\csname bibitem#1\endcsname}%
\let\auto@bib@innerbib\@empty
\bibitem [{\citenamefont {Adler}(1994)}]{Adler:94}%
  \BibitemOpen
  \bibfield  {author} {\bibinfo {author} {\bibnamefont {Adler}, \bibfnamefont
  {S.~L.}}} (\bibinfo {year} {1994}),\ \href@noop {} {\bibfield  {journal}
  {\bibinfo  {journal} {Nucl. Phys. B}\ }\textbf {\bibinfo {volume} {415}},\
  \bibinfo {pages} {195}}\BibitemShut {NoStop}%
\bibitem [{\citenamefont {Adler}(2004)}]{Adler:04}%
  \BibitemOpen
  \bibfield  {author} {\bibinfo {author} {\bibnamefont {Adler}, \bibfnamefont
  {S.~L.}}} (\bibinfo {year} {2004}),\ \href@noop {} {\emph {\bibinfo {title}
  {Quantum theory as an emergent phenomenon}}}\ (\bibinfo  {publisher}
  {Cambridge University Press},\ \bibinfo {address} {Cambridge})\BibitemShut
  {NoStop}%
\bibitem [{\citenamefont {Adler}(2007)}]{Adler3:07}%
  \BibitemOpen
  \bibfield  {author} {\bibinfo {author} {\bibnamefont {Adler}, \bibfnamefont
  {S.~L.}}} (\bibinfo {year} {2007}),\ \href@noop {} {\bibfield  {journal}
  {\bibinfo  {journal} {J. Phys. A}\ }\textbf {\bibinfo {volume} {40}},\
  \bibinfo {pages} {2935}}\BibitemShut {NoStop}%
\bibitem [{\citenamefont {Adler}\ and\ \citenamefont
  {Millard}(1996)}]{Adler-Millard:1996}%
  \BibitemOpen
  \bibfield  {author} {\bibinfo {author} {\bibnamefont {Adler}, \bibfnamefont
  {S.~L.}}, \ and\ \bibinfo {author} {\bibfnamefont {A.~C.}\ \bibnamefont
  {Millard}}} (\bibinfo {year} {1996}),\ \href@noop {} {\bibfield  {journal}
  {\bibinfo  {journal} {Nucl. Phys. B}\ }\textbf {\bibinfo {volume} {473}},\
  \bibinfo {pages} {199}}\BibitemShut {NoStop}%
\bibitem [{\citenamefont {Arndt}\ and\ \citenamefont
  {Hornberger}(2014)}]{Arndt:2014}%
  \BibitemOpen
  \bibfield  {author} {\bibinfo {author} {\bibnamefont {Arndt}, \bibfnamefont
  {M.}}, \ and\ \bibinfo {author} {\bibfnamefont {K.}~\bibnamefont
  {Hornberger}}} (\bibinfo {year} {2014}),\ \href@noop {} {\bibfield  {journal}
  {\bibinfo  {journal} {Nature Physics}\ }\textbf {\bibinfo {volume} {10}},\
  \bibinfo {pages} {271}}\BibitemShut {NoStop}%
\bibitem [{\citenamefont {Aspelmeyer}\ \emph {et~al.}(2014)\citenamefont
  {Aspelmeyer}, \citenamefont {Kippenberg},\ and\ \citenamefont
  {Marquardt}}]{Aspelmeyer:2014}%
  \BibitemOpen
  \bibfield  {author} {\bibinfo {author} {\bibnamefont {Aspelmeyer},
  \bibfnamefont {M.}}, \bibinfo {author} {\bibfnamefont {T.~J.}\ \bibnamefont
  {Kippenberg}}, \ and\ \bibinfo {author} {\bibfnamefont {F.}~\bibnamefont
  {Marquardt}}} (\bibinfo {year} {2014}),\ \href@noop {} {\bibfield  {journal}
  {\bibinfo  {journal} {Rev. Mod. Phys.}\ }\textbf {\bibinfo {volume} {86}},\
  \bibinfo {pages} {1391}}\BibitemShut {NoStop}%
\bibitem [{\citenamefont {Bahrami}\ \emph
  {et~al.}(2014{\natexlab{a}})\citenamefont {Bahrami}, \citenamefont {Bassi},\
  and\ \citenamefont {Ulbricht}}]{Bahrami:2014}%
  \BibitemOpen
  \bibfield  {author} {\bibinfo {author} {\bibnamefont {Bahrami}, \bibfnamefont
  {M.}}, \bibinfo {author} {\bibfnamefont {A.}~\bibnamefont {Bassi}}, \ and\
  \bibinfo {author} {\bibfnamefont {H.}~\bibnamefont {Ulbricht}}} (\bibinfo
  {year} {2014}{\natexlab{a}}),\ \href@noop {} {\bibfield  {journal} {\bibinfo
  {journal} {Phys. Rev. A}\ }\textbf {\bibinfo {volume} {89}},\ \bibinfo
  {pages} {032127}}\BibitemShut {NoStop}%
\bibitem [{\citenamefont {Bahrami}\ \emph
  {et~al.}(2014{\natexlab{b}})\citenamefont {Bahrami}, \citenamefont
  {Paternostro}, \citenamefont {Bassi},\ and\ \citenamefont
  {Ulbricht}}]{Bahrami:2014a}%
  \BibitemOpen
  \bibfield  {author} {\bibinfo {author} {\bibnamefont {Bahrami}, \bibfnamefont
  {M.}}, \bibinfo {author} {\bibfnamefont {M.}~\bibnamefont {Paternostro}},
  \bibinfo {author} {\bibfnamefont {A.}~\bibnamefont {Bassi}}, \ and\ \bibinfo
  {author} {\bibfnamefont {H.}~\bibnamefont {Ulbricht}}} (\bibinfo {year}
  {2014}{\natexlab{b}}),\ \href@noop {} {\bibfield  {journal} {\bibinfo
  {journal} {Phys. Rev. Lett.}\ }\textbf {\bibinfo {volume} {112}},\ \bibinfo
  {pages} {210404}}\BibitemShut {NoStop}%
\bibitem [{\citenamefont {Banerjee}\ \emph {et~al.}(2016)\citenamefont
  {Banerjee}, \citenamefont {Bera},\ and\ \citenamefont
  {Singh}}]{Banerjee:2016}%
  \BibitemOpen
  \bibfield  {author} {\bibinfo {author} {\bibnamefont {Banerjee},
  \bibfnamefont {S.}}, \bibinfo {author} {\bibfnamefont {S.}~\bibnamefont
  {Bera}}, \ and\ \bibinfo {author} {\bibfnamefont {T.~P.}\ \bibnamefont
  {Singh}}} (\bibinfo {year} {2016}),\ \href@noop {} {\bibfield  {journal}
  {\bibinfo  {journal} {Int. J. Mod. Phys.}\ }\textbf {\bibinfo {volume}
  {25}},\ \bibinfo {pages} {1644005}}\BibitemShut {NoStop}%
\bibitem [{\citenamefont {Bassi}\ and\ \citenamefont
  {Ghirardi}(2003)}]{Bassi:03}%
  \BibitemOpen
  \bibfield  {author} {\bibinfo {author} {\bibnamefont {Bassi}, \bibfnamefont
  {A.}}, \ and\ \bibinfo {author} {\bibfnamefont {G.~C.}\ \bibnamefont
  {Ghirardi}}} (\bibinfo {year} {2003}),\ \href@noop {} {\bibfield  {journal}
  {\bibinfo  {journal} {Phys. Rep.}\ }\textbf {\bibinfo {volume} {379}},\
  \bibinfo {pages} {257}}\BibitemShut {NoStop}%
\bibitem [{\citenamefont {Bassi}\ \emph {et~al.}(2013)\citenamefont {Bassi},
  \citenamefont {Lochan}, \citenamefont {Satin}, \citenamefont {Singh},\ and\
  \citenamefont {Ulbricht}}]{RMP:2012}%
  \BibitemOpen
  \bibfield  {author} {\bibinfo {author} {\bibnamefont {Bassi}, \bibfnamefont
  {A.}}, \bibinfo {author} {\bibfnamefont {K.}~\bibnamefont {Lochan}}, \bibinfo
  {author} {\bibfnamefont {S.}~\bibnamefont {Satin}}, \bibinfo {author}
  {\bibfnamefont {T.~P.}\ \bibnamefont {Singh}}, \ and\ \bibinfo {author}
  {\bibfnamefont {H.}~\bibnamefont {Ulbricht}}} (\bibinfo {year} {2013}),\
  \href@noop {} {\bibfield  {journal} {\bibinfo  {journal} {Rev. Mod. Phys.}\
  }\textbf {\bibinfo {volume} {85}},\ \bibinfo {pages} {471}}\BibitemShut
  {NoStop}%
\bibitem [{\citenamefont {Bedingham}(2016)}]{Bedingham:2016a}%
  \BibitemOpen
  \bibfield  {author} {\bibinfo {author} {\bibnamefont {Bedingham},
  \bibfnamefont {D.}}} (\bibinfo {year} {2016}),\ \href@noop {} {\ \textbf
  {\bibinfo {volume} {arXiv:1612.09470 [quant-ph]}}}\BibitemShut {NoStop}%
\bibitem [{\citenamefont {Bedingham}(2011{\natexlab{a}})}]{Bedingham:10}%
  \BibitemOpen
  \bibfield  {author} {\bibinfo {author} {\bibnamefont {Bedingham},
  \bibfnamefont {D.~J.}}} (\bibinfo {year} {2011}{\natexlab{a}}),\ \href@noop
  {} {\bibfield  {journal} {\bibinfo  {journal} {Found. Phys.}\ }\textbf
  {\bibinfo {volume} {41}},\ \bibinfo {pages} {686
  [arXiv:1003.2774]}}\BibitemShut {NoStop}%
\bibitem [{\citenamefont {Bedingham}(2011{\natexlab{b}})}]{Bedingham:11}%
  \BibitemOpen
  \bibfield  {author} {\bibinfo {author} {\bibnamefont {Bedingham},
  \bibfnamefont {D.~J.}}} (\bibinfo {year} {2011}{\natexlab{b}}),\ \href@noop
  {} {\ \textbf {\bibinfo {volume} {arXiv:1103.3974}}}\BibitemShut {NoStop}%
\bibitem [{\citenamefont {Bera}\ \emph {et~al.}(2015)\citenamefont {Bera},
  \citenamefont {Motwani}, \citenamefont {Singh},\ and\ \citenamefont
  {Ulbricht}}]{Bera:2015}%
  \BibitemOpen
  \bibfield  {author} {\bibinfo {author} {\bibnamefont {Bera}, \bibfnamefont
  {S.}}, \bibinfo {author} {\bibfnamefont {B.}~\bibnamefont {Motwani}},
  \bibinfo {author} {\bibfnamefont {T.~P.}\ \bibnamefont {Singh}}, \ and\
  \bibinfo {author} {\bibfnamefont {H.}~\bibnamefont {Ulbricht}}} (\bibinfo
  {year} {2015}),\ \href@noop {} {\bibfield  {journal} {\bibinfo  {journal}
  {Scientific Reports}\ }\textbf {\bibinfo {volume} {5}},\ \bibinfo {pages}
  {7664}}\BibitemShut {NoStop}%
\bibitem [{\citenamefont {Billardello}\ \emph {et~al.}(2016)\citenamefont
  {Billardello}, \citenamefont {Trombettoni},\ and\ \citenamefont
  {Bassi}}]{Bilardello:2016}%
  \BibitemOpen
  \bibfield  {author} {\bibinfo {author} {\bibnamefont {Billardello},
  \bibfnamefont {M.}}, \bibinfo {author} {\bibfnamefont {A.}~\bibnamefont
  {Trombettoni}}, \ and\ \bibinfo {author} {\bibfnamefont {A.}~\bibnamefont
  {Bassi}}} (\bibinfo {year} {2016}),\ \href@noop {} {\ \textbf {\bibinfo
  {volume} {arXiv:1612.0769 [quant-ph]}}}\BibitemShut {NoStop}%
\bibitem [{\citenamefont {Carlesso}\ \emph {et~al.}(2016)\citenamefont
  {Carlesso}, \citenamefont {Bassi}, \citenamefont {Falferi},\ and\
  \citenamefont {Vinante}}]{Carlesso:2016}%
  \BibitemOpen
  \bibfield  {author} {\bibinfo {author} {\bibnamefont {Carlesso},
  \bibfnamefont {M.}}, \bibinfo {author} {\bibfnamefont {A.}~\bibnamefont
  {Bassi}}, \bibinfo {author} {\bibfnamefont {P.}~\bibnamefont {Falferi}}, \
  and\ \bibinfo {author} {\bibfnamefont {A.}~\bibnamefont {Vinante}}} (\bibinfo
  {year} {2016}),\ \href@noop {} {\bibfield  {journal} {\bibinfo  {journal}
  {Phys. Rev. D}\ }\textbf {\bibinfo {volume} {94}},\ \bibinfo {pages}
  {124036}}\BibitemShut {NoStop}%
\bibitem [{\citenamefont {Collett}\ and\ \citenamefont
  {Pearle}(2003)}]{PEARLE5}%
  \BibitemOpen
  \bibfield  {author} {\bibinfo {author} {\bibnamefont {Collett}, \bibfnamefont
  {B.}}, \ and\ \bibinfo {author} {\bibfnamefont {P.}~\bibnamefont {Pearle}}}
  (\bibinfo {year} {2003}),\ \href@noop {} {\bibfield  {journal} {\bibinfo
  {journal} {Found. Phys.}\ }\textbf {\bibinfo {volume} {33}},\ \bibinfo
  {pages} {1495}}\BibitemShut {NoStop}%
\bibitem [{\citenamefont {Dowker}\ and\ \citenamefont
  {Henson}(2004)}]{Dowker:04}%
  \BibitemOpen
  \bibfield  {author} {\bibinfo {author} {\bibnamefont {Dowker}, \bibfnamefont
  {F.}}, \ and\ \bibinfo {author} {\bibfnamefont {J.}~\bibnamefont {Henson}}}
  (\bibinfo {year} {2004}),\ \href@noop {} {\bibfield  {journal} {\bibinfo
  {journal} {Journal of Statistical Physics}\ }\textbf {\bibinfo {volume}
  {115}},\ \bibinfo {pages} {1327}}\BibitemShut {NoStop}%
\bibitem [{\citenamefont {Dowker}\ and\ \citenamefont
  {Herbauts}(2004)}]{Dowker2:04}%
  \BibitemOpen
  \bibfield  {author} {\bibinfo {author} {\bibnamefont {Dowker}, \bibfnamefont
  {F.}}, \ and\ \bibinfo {author} {\bibfnamefont {I.}~\bibnamefont {Herbauts}}}
  (\bibinfo {year} {2004}),\ \href@noop {} {\bibfield  {journal} {\bibinfo
  {journal} {Classical and Quantum Gravity}\ }\textbf {\bibinfo {volume}
  {21}},\ \bibinfo {pages} {2963}}\BibitemShut {NoStop}%
\bibitem [{\citenamefont {Ghirardi}\ \emph
  {et~al.}(1990{\natexlab{a}})\citenamefont {Ghirardi}, \citenamefont
  {Grassi},\ and\ \citenamefont {Pearle}}]{Ghirardi:90}%
  \BibitemOpen
  \bibfield  {author} {\bibinfo {author} {\bibnamefont {Ghirardi},
  \bibfnamefont {G.~C.}}, \bibinfo {author} {\bibfnamefont {R.}~\bibnamefont
  {Grassi}}, \ and\ \bibinfo {author} {\bibfnamefont {P.}~\bibnamefont
  {Pearle}}} (\bibinfo {year} {1990}{\natexlab{a}}),\ \href@noop {} {\bibfield
  {journal} {\bibinfo  {journal} {Found. Phys.}\ }\textbf {\bibinfo {volume}
  {20}},\ \bibinfo {pages} {1271}}\BibitemShut {NoStop}%
\bibitem [{\citenamefont {Ghirardi}\ \emph
  {et~al.}(1990{\natexlab{b}})\citenamefont {Ghirardi}, \citenamefont
  {Pearle},\ and\ \citenamefont {Rimini}}]{Ghirardi2:90}%
  \BibitemOpen
  \bibfield  {author} {\bibinfo {author} {\bibnamefont {Ghirardi},
  \bibfnamefont {G.~C.}}, \bibinfo {author} {\bibfnamefont {P.}~\bibnamefont
  {Pearle}}, \ and\ \bibinfo {author} {\bibfnamefont {A.}~\bibnamefont
  {Rimini}}} (\bibinfo {year} {1990}{\natexlab{b}}),\ \href@noop {} {\bibfield
  {journal} {\bibinfo  {journal} {Phys. Rev. A}\ }\textbf {\bibinfo {volume}
  {42}},\ \bibinfo {pages} {78}}\BibitemShut {NoStop}%
\bibitem [{\citenamefont {Ghirardi}\ \emph {et~al.}(1986)\citenamefont
  {Ghirardi}, \citenamefont {Rimini},\ and\ \citenamefont
  {Weber}}]{Ghirardi:86}%
  \BibitemOpen
  \bibfield  {author} {\bibinfo {author} {\bibnamefont {Ghirardi},
  \bibfnamefont {G.~C.}}, \bibinfo {author} {\bibfnamefont {A.}~\bibnamefont
  {Rimini}}, \ and\ \bibinfo {author} {\bibfnamefont {T.}~\bibnamefont
  {Weber}}} (\bibinfo {year} {1986}),\ \href@noop {} {\bibfield  {journal}
  {\bibinfo  {journal} {Phys. Rev. D}\ }\textbf {\bibinfo {volume} {34}},\
  \bibinfo {pages} {470}}\BibitemShut {NoStop}%
\bibitem [{\citenamefont {Goldwater}\ \emph {et~al.}(2015)\citenamefont
  {Goldwater}, \citenamefont {Paternostro},\ and\ \citenamefont
  {Barker}}]{Goldwater:2015}%
  \BibitemOpen
  \bibfield  {author} {\bibinfo {author} {\bibnamefont {Goldwater},
  \bibfnamefont {D.}}, \bibinfo {author} {\bibfnamefont {M.}~\bibnamefont
  {Paternostro}}, \ and\ \bibinfo {author} {\bibfnamefont {P.}~\bibnamefont
  {Barker}}} (\bibinfo {year} {2015}),\ \href@noop {} {\bibfield  {journal}
  {\bibinfo  {journal} {Phys. Rev. A}\ }\textbf {\bibinfo {volume} {94}},\
  \bibinfo {pages} {010104}}\BibitemShut {NoStop}%
\bibitem [{\citenamefont {Laloe}\ \emph {et~al.}(2014)\citenamefont {Laloe},
  \citenamefont {Mullin},\ and\ \citenamefont {Pearle}}]{Laloe:2014}%
  \BibitemOpen
  \bibfield  {author} {\bibinfo {author} {\bibnamefont {Laloe}, \bibfnamefont
  {F.}}, \bibinfo {author} {\bibfnamefont {W.~J.}\ \bibnamefont {Mullin}}, \
  and\ \bibinfo {author} {\bibfnamefont {P.}~\bibnamefont {Pearle}}} (\bibinfo
  {year} {2014}),\ \href@noop {} {\bibfield  {journal} {\bibinfo  {journal}
  {Phys. Rev. A}\ }\textbf {\bibinfo {volume} {90}},\ \bibinfo {pages}
  {052119}}\BibitemShut {NoStop}%
\bibitem [{\citenamefont {Li}\ \emph {et~al.}(2016)\citenamefont {Li},
  \citenamefont {Steane}, \citenamefont {Bedingham},\ and\ \citenamefont
  {Briggs}}]{Bedingham:2016}%
  \BibitemOpen
  \bibfield  {author} {\bibinfo {author} {\bibnamefont {Li}, \bibfnamefont
  {Y.}}, \bibinfo {author} {\bibfnamefont {A.~M.}\ \bibnamefont {Steane}},
  \bibinfo {author} {\bibfnamefont {D.}~\bibnamefont {Bedingham}}, \ and\
  \bibinfo {author} {\bibfnamefont {G.~A.~D.}\ \bibnamefont {Briggs}}}
  (\bibinfo {year} {2016}),\ \href@noop {} {\ \textbf {\bibinfo {volume}
  {arXiv:1605.01881 [quant-ph]}}}\BibitemShut {NoStop}%
\bibitem [{\citenamefont {Lochan}\ \emph {et~al.}(2012)\citenamefont {Lochan},
  \citenamefont {Satin},\ and\ \citenamefont {Singh}}]{Lochan:2012}%
  \BibitemOpen
  \bibfield  {author} {\bibinfo {author} {\bibnamefont {Lochan}, \bibfnamefont
  {K.}}, \bibinfo {author} {\bibfnamefont {S.}~\bibnamefont {Satin}}, \ and\
  \bibinfo {author} {\bibfnamefont {T.~P.}\ \bibnamefont {Singh}}} (\bibinfo
  {year} {2012}),\ \href@noop {} {\bibfield  {journal} {\bibinfo  {journal}
  {Found. Phys.}\ }\textbf {\bibinfo {volume} {42}},\ \bibinfo {pages}
  {1556}}\BibitemShut {NoStop}%
\bibitem [{\citenamefont {Lochan}\ and\ \citenamefont
  {Singh}(2011)}]{Lochan-Singh:2011}%
  \BibitemOpen
  \bibfield  {author} {\bibinfo {author} {\bibnamefont {Lochan}, \bibfnamefont
  {K.}}, \ and\ \bibinfo {author} {\bibfnamefont {T.~P.}\ \bibnamefont
  {Singh}}} (\bibinfo {year} {2011}),\ \href@noop {} {\bibfield  {journal}
  {\bibinfo  {journal} {Phys. Lett. A}\ }\textbf {\bibinfo {volume} {375}},\
  \bibinfo {pages} {3747}}\BibitemShut {NoStop}%
\bibitem [{\citenamefont {Millen}\ \emph {et~al.}(2015)\citenamefont {Millen},
  \citenamefont {Fonesca}, \citenamefont {Mavrogordatos}, \citenamefont
  {Monteiro},\ and\ \citenamefont {Barker}}]{Barker:2015}%
  \BibitemOpen
  \bibfield  {author} {\bibinfo {author} {\bibnamefont {Millen}, \bibfnamefont
  {J.}}, \bibinfo {author} {\bibfnamefont {P.~Z.~G.}\ \bibnamefont {Fonesca}},
  \bibinfo {author} {\bibfnamefont {T.}~\bibnamefont {Mavrogordatos}}, \bibinfo
  {author} {\bibfnamefont {T.~S.}\ \bibnamefont {Monteiro}}, \ and\ \bibinfo
  {author} {\bibfnamefont {P.}~\bibnamefont {Barker}}} (\bibinfo {year}
  {2015}),\ \href@noop {} {\bibfield  {journal} {\bibinfo  {journal} {Phys.
  Rev. Lett.}\ }\textbf {\bibinfo {volume} {114}},\ \bibinfo {pages}
  {123602}}\BibitemShut {NoStop}%
\bibitem [{\citenamefont {Myrvold}\ \emph {et~al.}(2009)\citenamefont
  {Myrvold}, \citenamefont {Christian},\ and\ \citenamefont
  {Pearle}}]{Myrvold:09}%
  \BibitemOpen
  \bibfield  {author} {\bibinfo {author} {\bibnamefont {Myrvold}, \bibfnamefont
  {W.~C.}}, \bibinfo {author} {\bibfnamefont {J.}~\bibnamefont {Christian}}, \
  and\ \bibinfo {author} {\bibfnamefont {P.}~\bibnamefont {Pearle}}} (\bibinfo
  {year} {2009}),\ in\ \href@noop {} {\emph {\bibinfo {booktitle} {Quantum
  Reality, Relativistic Causality, and Closing the Epistemic Circle}}},\
  \bibinfo {series} {The Western Ontario Series in Philosophy of Science},
  Vol.~\bibinfo {volume} {73},\ \bibinfo {editor} {edited by\ \bibinfo {editor}
  {\bibfnamefont {W.}~\bibnamefont {Demopoulos}}, \bibinfo {editor}
  {\bibfnamefont {D.}~\bibnamefont {Devidi}}, \bibinfo {editor} {\bibfnamefont
  {R.}~\bibnamefont {Disalle}}, \ and\ \bibinfo {editor} {\bibfnamefont
  {W.}~\bibnamefont {Myrvold}}}\ (\bibinfo  {publisher} {Springer
  Netherlands})\ pp.\ \bibinfo {pages} {257--292}\BibitemShut {NoStop}%
\bibitem [{\citenamefont {Nicrosini}\ and\ \citenamefont
  {Rimini}(2003)}]{Nicrosini:03}%
  \BibitemOpen
  \bibfield  {author} {\bibinfo {author} {\bibnamefont {Nicrosini},
  \bibfnamefont {O.}}, \ and\ \bibinfo {author} {\bibfnamefont
  {A.}~\bibnamefont {Rimini}}} (\bibinfo {year} {2003}),\ \href@noop {}
  {\bibfield  {journal} {\bibinfo  {journal} {Foundations of Physics}\ }\textbf
  {\bibinfo {volume} {33}},\ \bibinfo {pages} {1061}}\BibitemShut {NoStop}%
\bibitem [{\citenamefont {Pearle}(1989)}]{Pearle:89}%
  \BibitemOpen
  \bibfield  {author} {\bibinfo {author} {\bibnamefont {Pearle}, \bibfnamefont
  {P.}}} (\bibinfo {year} {1989}),\ \href@noop {} {\bibfield  {journal}
  {\bibinfo  {journal} {Phys. Rev. A}\ }\textbf {\bibinfo {volume} {39}},\
  \bibinfo {pages} {2277}}\BibitemShut {NoStop}%
\bibitem [{\citenamefont {Pearle}(1990)}]{Pearle:90}%
  \BibitemOpen
  \bibfield  {author} {\bibinfo {author} {\bibnamefont {Pearle}, \bibfnamefont
  {P.}}} (\bibinfo {year} {1990}),\ in\ \href@noop {} {\emph {\bibinfo
  {booktitle} {Sixty-Two Years of Uncertainity: Historical Philosophical, and
  Physics Inquires into the Foundations of Quantum Physics}}},\ \bibinfo
  {editor} {edited by\ \bibinfo {editor} {\bibfnamefont {A.~I.}\ \bibnamefont
  {Miller}}}\ (\bibinfo  {publisher} {Plenum, New York})\BibitemShut {NoStop}%
\bibitem [{\citenamefont {Pearle}(1999)}]{Pearle2:99}%
  \BibitemOpen
  \bibfield  {author} {\bibinfo {author} {\bibnamefont {Pearle}, \bibfnamefont
  {P.}}} (\bibinfo {year} {1999}),\ \href {\doibase 10.1103/PhysRevA.59.80}
  {\bibfield  {journal} {\bibinfo  {journal} {Phys. Rev. A}\ }\textbf {\bibinfo
  {volume} {59}},\ \bibinfo {pages} {80}}\BibitemShut {NoStop}%
\bibitem [{\citenamefont {Pearle}\ and\ \citenamefont
  {Squires}(1994)}]{Pearle:94}%
  \BibitemOpen
  \bibfield  {author} {\bibinfo {author} {\bibnamefont {Pearle}, \bibfnamefont
  {P.}}, \ and\ \bibinfo {author} {\bibfnamefont {E.}~\bibnamefont {Squires}}}
  (\bibinfo {year} {1994}),\ \href {\doibase 10.1103/PhysRevLett.73.1}
  {\bibfield  {journal} {\bibinfo  {journal} {Phys. Rev. Lett.}\ }\textbf
  {\bibinfo {volume} {73}},\ \bibinfo {pages} {1}}\BibitemShut {NoStop}%
\bibitem [{\citenamefont {Rashid}\ \emph {et~al.}(2016)\citenamefont {Rashid},
  \citenamefont {Tommaso}, \citenamefont {Bateman}, \citenamefont {Vovrosh},
  \citenamefont {Hempston}, \citenamefont {Kim},\ and\ \citenamefont
  {H}}]{Rashid:2016}%
  \BibitemOpen
  \bibfield  {author} {\bibinfo {author} {\bibnamefont {Rashid}, \bibfnamefont
  {M.}}, \bibinfo {author} {\bibfnamefont {T.}~\bibnamefont {Tommaso}},
  \bibinfo {author} {\bibfnamefont {J.}~\bibnamefont {Bateman}}, \bibinfo
  {author} {\bibfnamefont {J.}~\bibnamefont {Vovrosh}}, \bibinfo {author}
  {\bibfnamefont {D.}~\bibnamefont {Hempston}}, \bibinfo {author}
  {\bibfnamefont {M.~S.}\ \bibnamefont {Kim}}, \ and\ \bibinfo {author}
  {\bibfnamefont {U.}~\bibnamefont {H}}} (\bibinfo {year} {2016}),\ \href@noop
  {} {\bibfield  {journal} {\bibinfo  {journal} {Phys. Rev. Lett.}\ }\textbf
  {\bibinfo {volume} {117}},\ \bibinfo {pages} {273601}}\BibitemShut {NoStop}%
\bibitem [{\citenamefont {Singh}(2006)}]{Singh:2006}%
  \BibitemOpen
  \bibfield  {author} {\bibinfo {author} {\bibnamefont {Singh}, \bibfnamefont
  {T.~P.}}} (\bibinfo {year} {2006}),\ \href@noop {} {\bibfield  {journal}
  {\bibinfo  {journal} {Bulg. J. Phys.}\ }\textbf {\bibinfo {volume} {33}},\
  \bibinfo {pages} {217}}\BibitemShut {NoStop}%
\bibitem [{\citenamefont {Singh}(2015)}]{Singh:2012}%
  \BibitemOpen
  \bibfield  {author} {\bibinfo {author} {\bibnamefont {Singh}, \bibfnamefont
  {T.~P.}}} (\bibinfo {year} {2015}),\ in\ \href@noop {} {\emph {\bibinfo
  {booktitle} {Re-thinking time at the interface of physics and philosophy: the
  forgotten present}}},\ \bibinfo {series and number} {(arXiv:1210.81110)},\
  \bibinfo {editor} {edited by\ \bibinfo {editor} {\bibfnamefont
  {T.}~\bibnamefont {Filk}}\ and\ \bibinfo {editor} {\bibfnamefont
  {A.}~\bibnamefont {von Muller}}}\ (\bibinfo  {publisher} {Berlin-Heidelberg :
  Springer})\BibitemShut {NoStop}%
\bibitem [{\citenamefont {Tumulka}(2006)}]{Tumulka2:06}%
  \BibitemOpen
  \bibfield  {author} {\bibinfo {author} {\bibnamefont {Tumulka}, \bibfnamefont
  {R.}}} (\bibinfo {year} {2006}),\ \href@noop {} {\bibfield  {journal}
  {\bibinfo  {journal} {Journal of Statistical Physics}\ }\textbf {\bibinfo
  {volume} {125}},\ \bibinfo {pages} {821}}\BibitemShut {NoStop}%
\bibitem [{\citenamefont {Vinante}\ \emph
  {et~al.}(2016{\natexlab{a}})\citenamefont {Vinante}, \citenamefont {Bahrami},
  \citenamefont {Bassi}, \citenamefont {Usenko}, \citenamefont {Wijts},\ and\
  \citenamefont {Oosterkamp}}]{Vinante:2016a}%
  \BibitemOpen
  \bibfield  {author} {\bibinfo {author} {\bibnamefont {Vinante}, \bibfnamefont
  {A.}}, \bibinfo {author} {\bibfnamefont {M.}~\bibnamefont {Bahrami}},
  \bibinfo {author} {\bibfnamefont {A.}~\bibnamefont {Bassi}}, \bibinfo
  {author} {\bibfnamefont {O.}~\bibnamefont {Usenko}}, \bibinfo {author}
  {\bibfnamefont {G.}~\bibnamefont {Wijts}}, \ and\ \bibinfo {author}
  {\bibfnamefont {T.~H.}\ \bibnamefont {Oosterkamp}}} (\bibinfo {year}
  {2016}{\natexlab{a}}),\ \href@noop {} {\bibfield  {journal} {\bibinfo
  {journal} {Phys. Rev. Lett.}\ }\textbf {\bibinfo {volume} {116}},\ \bibinfo
  {pages} {090402}}\BibitemShut {NoStop}%
\bibitem [{\citenamefont {Vinante}\ \emph
  {et~al.}(2016{\natexlab{b}})\citenamefont {Vinante}, \citenamefont
  {Mezzena},\ and\ \citenamefont {Falferi}}]{Vinante:2016b}%
  \BibitemOpen
  \bibfield  {author} {\bibinfo {author} {\bibnamefont {Vinante}, \bibfnamefont
  {A.}}, \bibinfo {author} {\bibfnamefont {R.}~\bibnamefont {Mezzena}}, \ and\
  \bibinfo {author} {\bibfnamefont {P.}~\bibnamefont {Falferi}}} (\bibinfo
  {year} {2016}{\natexlab{b}}),\ \href@noop {} {\ \textbf {\bibinfo {volume}
  {arXiv:1611.09776 [quant-ph]}}}\BibitemShut {NoStop}%
\bibitem [{\citenamefont {Wan}\ \emph {et~al.}(2016)\citenamefont {Wan},
  \citenamefont {Scala}, \citenamefont {Morley}, \citenamefont {Rahman},
  \citenamefont {Ulbricht}, \citenamefont {Bateman}, \citenamefont {Bose},\
  and\ \citenamefont {Kim}}]{Ulbricht:2016}%
  \BibitemOpen
  \bibfield  {author} {\bibinfo {author} {\bibnamefont {Wan}, \bibfnamefont
  {C.}}, \bibinfo {author} {\bibfnamefont {M.}~\bibnamefont {Scala}}, \bibinfo
  {author} {\bibfnamefont {G.}~\bibnamefont {Morley}}, \bibinfo {author}
  {\bibfnamefont {A.}~\bibnamefont {Rahman}}, \bibinfo {author} {\bibnamefont
  {Ulbricht}}, \bibinfo {author} {\bibfnamefont {P.~F.}\ \bibnamefont
  {Bateman}, \bibfnamefont {H.~J.and~Barker}}, \bibinfo {author} {\bibfnamefont
  {S.}~\bibnamefont {Bose}}, \ and\ \bibinfo {author} {\bibfnamefont
  {M.}~\bibnamefont {Kim}}} (\bibinfo {year} {2016}),\ \href@noop {} {\bibfield
   {journal} {\bibinfo  {journal} {Phys. Rev. Lett.}\ }\textbf {\bibinfo
  {volume} {117}},\ \bibinfo {pages} {143003}}\BibitemShut {NoStop}%
\end{thebibliography}%

\end{document}